# Addressing contingency in algorithmic (mis)information classification: Toward a responsible machine learning agenda

Andrés Domínguez Hernández*
Department of Computer Science, University of Bristol, andres.dominguez@bristol.ac.uk

Richard Owen
School of Management, University of Bristol

Dan Saattrup Nielsen
Department of Engineering Mathematics, University of Bristol

Ryan McConville
Department of Engineering Mathematics, University of Bristol
*Corresponding author

## Abstract

Machine learning (ML) enabled classification models are becoming increasingly popular for tackling the sheer volume and speed of online misinformation and other content that could be identified as harmful. In building these models, data scientists need to take a stance on the legitimacy, authoritativeness and objectivity of the sources of "truth" used for model training and testing. This has political, ethical and epistemic implications which are rarely addressed in technical papers. Despite (and due to) their reported high accuracy and performance, ML-driven moderation systems have the potential to shape online public debate and create downstream negative impacts such as undue censorship and the reinforcing of false beliefs. Using collaborative ethnography and theoretical insights from social studies of science and expertise, we offer a critical analysis of the process of building ML models for (mis)information classification: we identify a series of algorithmic contingencies—key moments during model development that could lead to different future outcomes, uncertainty and harmful effects as these tools are deployed by social media platforms. We conclude by offering a tentative path toward reflexive and responsible development of ML tools for moderating misinformation and other harmful content online.

# 1 Introduction

In recent years there has been a flurry of research on the automated detection of misinformation using Machine Learning (ML) techniques. Significant progress has been made on developing ML models for the identification, early detection and management of online misinformation,[1] which can then be deployed at scale to assist human moderators, e.g., [34,50,77]. The development of these tools has gained currency particularly among social media platforms like Meta, Twitter and YouTube given their key role in the propagation of online misinformation and mounting regulatory pressure to manage the problem. In response to the overwhelming scale of misinformation –notably in the context of the COVID-19 pandemic- and the limited capacity of human moderation to address this, platforms have increasingly looked to the deployment of automated models as standalone solutions requiring less or no human intervention [9].

The artificial intelligence (AI) research community has broadly framed the problem as one that can be tackled using ML–enabled classification models. These classify, with varying levels of accuracy, the category to which a piece of data belongs (e.g. "factually true", "false" or "misleading" claim). These models are trained on large datasets of various modalities (images, text or social connections) containing manually annotated samples of information labelled as being factually correct or false [70]. In order to advance the state of the art, researchers strongly emphasize the need for more and higher quality data which can be used to train and validate ML models. Several training datasets have been published to this end containing collections of fake news articles, social media posts, fabricated images, or false claims along with labels about their truthfulness produced by fact-checking organizations around the world.[2]

Recent work in fair-ML and critical data studies has started to examine the assumptions and practices surrounding the curation of training sets and their use in the construction and application of ML models. Of note are discussions relating to ethical issues of algorithmic discrimination, bias and unfairness, e.g., [4,13,37,48]. However, scant attention has been given to the epistemic assumptions that underlie ML-enabled models for misinformation classification and, associated with this, their social, political and commercial entanglements –what we shall refer to as algorithmic contingencies. A persistent rationale in developing such tools is that if fed with (large and good) enough data they will be able to produce reliable, actionable evaluations of truthfulness, allowing users to tackle the problem of misinformation in an automated, cost-effective manner. The construction of referential datasets (or "ground truths") used for both model training and performance measuring purposes is rooted in assumptions about the credibility and trustworthiness of these data sources. These sources typically include corpora of authoritative knowledge or the outputs of professional fact-checking organizations, which are implicitly assumed to be credible and can be used to benchmark what is "true"—or at least not false. Explanations about what counts as "authoritative" or "reliable" ground truths and reflection on associated assumptions, limitations and ethical implications are rarely seen in technical papers describing model development and application.

This paper addresses this issue by proposing a tentative agenda for responsible ML-enabled (mis)information classification. We present findings from a collaborative study between science and technology studies (STS) scholars and data scientists developing ML-enabled tools to combat misinformation. The collaboration grew out of an effort to embed responsible innovation into technical projects funded under a large interdisciplinary research center. Drawing insights from feminist epistemology and social studies of science and expertise, we lay out a series of *algorithmic contingencies*—key moments during model development which could lead to different future outcomes, uncertainties and harmful downstream impacts. The frame of contingencies departs from the calculus of fairness or data bias elimination discussed in the literature to date [63]. We advance that taking these contingencies seriously opens a space for reflection and evaluation of the social value and potential harmful impacts of these tools.

---

[1] Several terms related to misinformation (e.g., disinformation and fake news) are used throughout this paper to refer to specific attempts in the literature to define and tackle related problems. However, the term misinformation, in its broadest sense, is preferred for analysis as it encompasses any type of misleading or false content presented as factual, regardless of intent.

[2] A non-exhaustive list of datasets is found on [18]

## 2   Theoretical lens: On the social construction of facts, facticity and fact-checking

Who gets to decide whether a conjecture or claim meets the quality and condition of being a fact—i.e., establishes its facticity—is a contentious and contested matter. Reducing the establishment of facticity to "checking" and "verifying" loses the richness and complexity of fact construction as an inherently social process [40]. As scholars within feminist epistemology, philosophy of science and social studies of science and expertise have compellingly argued, facts do not exist in a value-free vacuum: they are crafted, contested and pondered against competing claims as they move within social worlds. Facts are thus necessarily contingent to context, cultural norms, institutional structures and power relations [12,32,36,40]. Not only this, but over time the social and historical circumstances on which the construction of a fact depends can become opaque and lost; seemingly "free from the circumstances of its production" [40:103]. In this shifting and contingent knowledge arena, the legitimacy of those who warrant and assert claims and conjectures becomes key.

Legitimacy can be granted through credentialled expertise, reputation and social acceptance [76]. Relying on the authority of scientific expertise and reputable journalism could well be a socially acceptable means for determining what one might call "objective knowledge". Feminist scholars have however warned against the objectivist ideal of "science as neutral" as it paradoxically elides the forces that often shape knowledge production –Western, male, and elite dominated funding institutions, research priorities, special interest groups, etc. [32,33]. This observation is not a relativist attack on expertise and science as an institution, but a call to remain cautious about the often loose use of the language of "neutrality" and "objectivity"  [33,46]. We take this caution as our starting point to examine how assumptions about (scientific) knowledge, expertise and facticity might become encoded in AI techniques aimed at sorting out and managing (mis)information.

While there are different computational approaches to combat misinformation, in this article we focus primarily on efforts to leverage and automate the journalistic practice of fact-checking through ML-based techniques [34]. In the last decade, professional fact-checking has gained prominence for its role in promoting truth in public discourse, especially during times of elections and crises (e.g., wars and pandemics). To date, there exist hundreds of professional fact checkers around the world.[3] Modern data-driven fact-checking has increasingly been viewed as vital to tackle misinformation in the so called "post truth" era [8]. Social media platforms, and notably Facebook, have partnered with professional fact-checkers around the world to combat the widespread misinformation problem.[4] Not only are fact-checkers entrusted with the moderation of dubious pieces of information flagged as such by platforms' algorithms, but their verdicts are used to help train misinformation detection algorithms and ML models [9]. The output and credibility of professional fact-checking is usually taken at face value for these purposes. However, as a human activity, professional fact-checking is not immune to cognitive and selection biases, subjective and ideological preferences, errors, and (geo)political and commercial interests. Despite being presented as impartial and objective, fact-checkers too are political actors engaging in epistemic practices, i.e., establishing facticity and confronting lies (defined by them) in public discourse [29].

Fact-checking services have attracted some criticism over their methodologies due to, for example, accusations of skewed selection of topics, actors, and claims; and the use of ambiguous terminology [66,72]. For instance, fact-checking organizations often use vague, or borderline terminology (e.g., "mostly true", "mostly false") on the basis that claims are not always verifiable nor simply true or false. These issues manifest in myriad ways; for example as competing or contentious verdicts between fact-checkers or shifting assessments of claims over time, sometimes with serious consequences [45,52].

Deceitful content and tactics are always evolving, but also, what constitutes a seemingly stable fact at a given point may change over time, driven by public debate or the availability of new information [47]. The COVID-19 lab leak controversy is a case in point. For the most part of 2020 the claim that COVID-19 originated in a lab in Wuhan, China, was widely dismissed by Western media as a conspiracy and "fake news". Early in 2021, growing calls to take the hypothesis seriously triggered further investigations by the WHO and a swift change of narrative by fact-checkers and the media [69]. Amidst the controversy, Facebook automatically mislabeled a news article critical of the WHO as "misinformation", which was later corrected after complaints of censorship by the news outlet[5]. This episode shows that while well intentioned, the practice of fact-checking can also lead to ambivalences and false positives which could in turn be blindly reproduced by an algorithm.

---

[3] A database of global fact-checking websites has identified more than 300, https://reporterslab.org/fact-checking/
[4] See Facebook AI: "Here's how we're using AI to help detect misinformation". November 19, 2020, https://ai.facebook.com/blog/heres-how-were-using-ai-to-help-detect-misinformation/

Not only the online misinformation ecology evolves quickly, but the experiences and manifestations of misinformation differ vastly across cultures, idiosyncrasies, languages and political realities [57,62]. These ambiguities are not trivial for the design of interventions as research has shown that the publication of fact-checks can have uneven effects on different audiences depending on a person's beliefs or initial stance on the topic [55,75]. Furthermore, correction efforts could have the backfiring effect of reinforcing entrenched beliefs and the spread of misinformation due to the segregating dynamics of knowledge communities online formed around shared politics and identities [53] (see also section 5).

In pointing out the challenges involved with "establishing the truth" we do not seek to undermine the value of expertise and journalism in public discourse. Albeit inevitably partial and context-dependent, truth-seeking efforts such as fact-checking can still be of use in the fight against misinformation. However, we contend that these practices and their normative claims warrant reflection and scrutiny, particularly as they get scaled up and automated.

## 3 Empirics and methodology

This article is the result of an interdisciplinary collaboration between data scientists leading a research project (CLARITI) focused on the use of machine learning for tackling online misinformation, and social scientists affiliated with the field of science and technology studies (STS) (hereafter the research team). Our analysis is informed by regular meetings within the research team over a period of approximately 8 months and empirically grounded in the development of a ML system to detect online misinformation over that same period. The collaboration was led by the social scientists (ADH and RO) as part of their crosscutting work on embedding reflexivity and responsible innovation within a major Centre hosted by their university.[6] The technical project was conducted by a team of data scientists (DSN and RM) and comprised a multimodal machine learning based study of misinformation on social media and the development of the ML system for misinformation detection, labelling and management. One of the outcomes of the data science project was a misinformation dataset [51] which intends to capture the diverse ways in which misinformation manifests on social media. This dataset contains roughly 13,000 claims (of which 95% are labelled as misinformation) from 115 fact-checking organizations and, more than 20 million posts ("tweets") from the Twitter platform related to these claims. Aside from capturing a sizeable amount of the social media interactions associated with the claims, the dataset covers 41 languages and spans dozens of different events (e.g., COVID-19, Israel-Palestine conflicts) appearing on the platform over the course of a decade.

This work builds on the longstanding tradition in STS of opening the world of scientists and black-boxed technical systems to scrutiny through ethnographic accounts [10,40] and extends previous efforts to integrate social and ethical considerations into processes of research and development [20,61]. While ethnography has been the archetypical tool of STS theory and critique, in this study we explicitly adopt a collaborative version of the method by shifting from an ethnographer/informer arrangement toward a joint endeavor between social scientists and data scientists c.f. [21]. Collaborative ethnography can be viewed as "an approach to ethnography that *deliberately* and *explicitly* emphasizes collaboration at every point in the ethnographic process, without veiling it—from project conceptualization, to fieldwork, and, especially, through the writing process. Collaborative ethnography invites commentary from our consultants and seeks to make that commentary overtly part of the ethnographic text as it develops" [39:16]. Our aim with this approach is not only to enrich the process of interpretation from field observations and qualitative analysis of technical work into writing, but to advance collective reflection and the co-development of ethical and responsible practices.

The discussions were aimed primarily at developing a schematization of the process of development of the ML detection model and the curation of the training datasets used to support this (see next section). We focused on the technical development phase of the project while concurrently analyzing comparable works in the literature on automatic misinformation detection. Given the collaborative nature of this work, the analysis combines interpretative description and self-critical reflections co-produced by the research team. We approached this method iteratively by purposely surfacing the technical and epistemic assumptions and practices in ML model development for analysis. Subsequently, interpretative texts written by the social scientists were checked and expanded by the data scientists. We acknowledge the limitations of our method in producing generalizing claims which are reflective of our own concerns, experiences, the practices of a specific project and a limited subset of works in the literature. The next two sections describe the findings of our study.

---

[5] See https://unherd.com/thepost/facebook-censors-award-winning-journalist-for-criticising-the-who/
[6] The National Research Centre on Privacy, Harm Reduction and Adversarial Influence Online REPHRAIN.

## 4 Contingencies of automatic misinformation detection

In this section we describe the steps taken in the data science project to construct the automated misinformation model. We use this as a tangible way to critically examine the assumptions and practices involved in the development of ML tools for online misinformation detection and outline some salient contingencies related to their construction.

Below we first describe the generic model construction process adopted in the project which includes the steps of problem definition, choice of variables, curation of ground truth datasets, model validation and finally deployment (Figure 1). We then employ the notion of contingencies to interrogate the entanglements associated with the model, and the conditions that could alter the actual or claimed utility of a model and lead to deleterious consequences. Harmful impacts could for example include legitimate information being wrongly categorized as misinformation (false positives) and subsequently leading to unfair censorship; the amplification of objectionable or ambiguous truth assessments; or the reinforcing of false beliefs by failing to identify misinformation (false negatives). It is important to note that here we do not view biases as inherently negative; in fact, they could be necessary for the purposes of tackling misinformation. For instance, using expert sources such as scientists or reputable institutions to correct misinformation is a form of socially acceptable bias which may prove effective even though experts are fallible and may not always reach consensus on what is true of even what constitutes as being a fact [40]. Thus, this exercise is not aimed at debiasing or showing how to better locate facts and determine "truth"; in fact, it highlights the difficulties of doing so via manual or computational approaches. We illustrate the salience of various contingencies so that they are pondered reflexively in the development and audit of tools. In the following subsections we examine these contingencies in more detail, describing them first and then locating each more specifically in the context of the project's research and development.

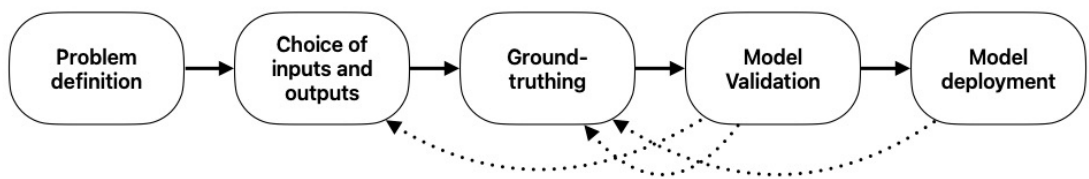

**Figure 1: Steps in the construction of a ML misinformation detection model. 1) Problem definition: design of strategy based on hypotheses, definitions and theories about how to identify misinformation. 2) Selection of multimodal inputs and outputs to be included into the classification model. 3) Ground-truthing: the ground truth dataset is used to train the model. A subset of the this is reserved for model validation. 4) Model validation: a subset of the ground truth dataset is used to test the model's performance. Metrics of performance accompany the publication of classification models. 5) Model deployment: model outputs inform online content moderation decisions such as banning, downranking or flagging.**

### 4.1 Problem definition

The way misinformation is problematized shapes the strategy used to detect and identify it as well as the required data required and their structure. Within the AI/ML research community, misinformation detection is generally framed as a task of *classification* whereby candidate pieces of information are classified by a ML model according to sensible (i.e.

legitimate) evaluations of their truthfulness.[7] These evaluations are influenced and informed by existing empirical studies, theories and hypotheses about what may constitute or signal the presence of misinformation. Misinformation and truth in this sense are defined a-priori by the researchers and then translated (*formalized*) into a ML model (see 4.2) following different data-driven strategies which could leverage e.g., a corpus of knowledge, writing content and style or patterns of propagation [27]. For example, "fake news"—a popularized idiom in the misinformation landscape—is typically defined as a type of misinformation which is intentionally crafted, often with a political or financial interest. Based on that hypothesis different indicators such as style of writing or other distinctive features could be leveraged to single out and identify fake news from the analyzed content (e.g., social media). For example, Rashkin et al. [58] developed a model that classifies political statements and news based on *linguistic features* such as keywords or subjective language that indicates signs of intent to deceive. In that case, the authors drew on previous empirical work, communication theory and hypotheses suggesting that "fake news articles try to enliven stories to attract readers". Techniques of natural language processing are increasingly being used to support this. Another common strategy is to search for signs of misinformation (irrespective of intent) by looking at its *impact,* particularly regarding what is distinctive about information consumption patterns in relation to those exhibited by legitimate, "truthful" information. For instance, several studies have shown that false information online (e.g., fake news) tends to spread faster than verified information [74]. A model informed by such findings would rely on the assumption that social media metrics—such as likes, retweets or comments—can reveal something about the patterns of consumption of falsehoods [50].

In the CLARITI project, multiple hypotheses underpinned the construction of the dataset which draw on previous studies, plausible assumptions or experiments conducted by the researchers. These were (1) people interact differently with posts discussing misinformation compared to posts discussing factually true information, in the sense of their replies and retweets [43,64]; (2) the images used when discussing misinformation are different to images used when discussing factually true information [38]; (3) users who are discussing misinformation tend to be different to those discussing factually true information as adjudged by their followers, followees and posts [16]; (4) misinformation spreads faster on social media than factually true claims [74]; (5) posts discussing misinformation tend to use different hashtags to posts discussing factually true claims, where a hashtag is assumed to be "different" from another hashtag depending on how it is used in tweets [14,44]; and (6) a classifier trained on data which is monolingual or monotopical will not generalize to new languages and events [31]. The variables related to these hypotheses were used as a basis to construct a dataset so as to make those variables available for analysis.

In sum, definitions, prior evidence and hypotheses (even if they are provisional) all inform further steps in the construction of the model and thereby become increasingly intractable and only auditable if explained sufficiently in a technical paper or documentation.

## 4.2 Choice of inputs and outputs

Once the problem is defined, it is formalised (or abstracted) into a mathematical function with inputs and outputs so that it can then be operationalised computationally. The choice of inputs to a ML model not only reflects the designer's framing of the problem but the technical feasibility in terms of what kind of data can be reliably and economically acquired and used at scale. Most algorithmic techniques to date have used text as an input variable – however misinformation is often also contained in images, video, or can result from deliberately mismatched combinations of text and images intended to deceive or lure users; hence the growing focus on working with multimodal data, which is also salient in the CLARITI project.

The lack of data is frequently mentioned in the literature as one of the biggest obstacles in the detection of misinformation. This is a twofold issue. On the one hand, some social networks make the acquiring of data more difficult (e.g., Meta) while others make this easier (e.g., Twitter where academic access is 10 million tweets per month, and it is straightforward to get access). On the other hand, in order to use supervised ML models, online content needs to be annotated e.g., as "true" or "false" (see 4.3). To get around the problem of general data scarcity, different forms of triangulation or proxy data are often included in the mix of building blocks. It may, for example, be relatively inexpensive to scrape social media for reactions or social engagement metrics which could be used to detect the presence of misinformation. For example, Lee et al., [41] use sentiment analysis to quantify the "perplexity" users express in their comments to social media posts as a proxy of early signs of misinformation. Because some models rely on statistical correlations between variables in lieu of causal relations, there is the risk of them making spurious associations. For instance, using data from satirical news as a source to identify fake news could lead to wrong associations due to the presence of confounding variables such as humor [56].

---

[7] Classification models in this domain are typically *supervised*, that is, they require properly annotated data for training and testing.

Opportunity, resources and feasibility were key considerations within the CLARITI project. The strategy was to use a *feature-rich* dataset containing as many modalities as possible with the expectation that the combination of these could lead to better (i.e. more accurate) predictions of misinformation. Given the problem was formulated as a binary classification task, the target variables were defined as "factual" and "misinformation". The model relied on annotated, multimodal data (i.e., texts and images) obtained from the Google Fact Check API[8] as well as from Twitter which provides access to data about user engagements with news. This was motivated by the ready availability of data for research purposes and because the widespread use of these sources in the automatic misinformation detection research literature would make it easier to test whether the current effort would *outperform* existing models.

## 4.3 Ground-truthing

In order to train a classification model, a labelled dataset –or ground truth—is needed as a referent of factual and non-factual information. This process of *ground-truthing* is not a trivial act but requires the machine learning developer to make principled choices as to what is an acceptable source of legitimate information [37]. As discussed above, a common and defensible approach to determining ground truths is to defer to experts or authoritative sources of knowledge. For example, Wikipedia, reputable news outlets, professional fact-checkers and "wisdom-of-the-crowd" have been used to build labelled datasets of categorized (mis)information. While using science-informed sources is seldom objectionable, there are still conditioning factors. For instance, in some circumstances, deference to experts may be unwarranted[9]; and journalists might (unintendedly or not) publish data in a way that is skewed and misleading [42]. For illustration, here we enumerate some of the conditionalities associated with ground truth datasets based on fact-checking.

First, ground truths are highly *contingent on timing* and thus *have diminishing returns*. This is because the online (mis)information environment is in constant flux. A model trained on previously fact-checked information is likely to be more effective with similar or comparable content and less so with whole new topics, themes and genres of false content. As the COVID-19 lab leak controversy demonstrates, factual assessments could shift dramatically over a short period of time. These shifts are not always adequately and consistently addressed by fact-checkers such that they can be taken onboard in updating an ML model.[10] If left unchecked, ambivalences in the training data could lead to the amplification of objectionable results and potentially harmful false positives.

Another key conditionality of constructing ground truth datasets is the choice of labels and labelling systems. This is particularly problematic when truth assessments are expressed in ways that are *ambiguous or subject to multiple interpretations*. One of the biggest challenges with using the work of fact-checkers as a source of ground truths is the lack of consistency among fact-checkers" definitions, terminology and methodology, particularly in cases where misinformation is not blatant, but subtle and nuanced. Different organizations use different types of labelling, including politically charged phrasing ("pants on fire"), borderline ("mostly true", "mostly false") or detailed assessments of claims when it comes to nuanced content which cannot be easily classified as either true or false.

---

[8] This service aggregates claims which have been fact-checked by eligible news organisations. To be included in Google's fact check tool, news organizations need to comply with Google's standards for publishers and content policies. https://support.google.com/news/publisher-center/answer/6204050

[9] An illustration of this is what Rietdijk and Archer [59] problematize as "false balance" in journalism. This issue has been particularly salient in the debate around climate change, where journalists have given disproportionate attention to a minority of climate sceptics within the scientific community, who may also qualify as experts, in their efforts to show both sides of the debate.

[10] For example, PolitiFact, a well-known fact-checker organization, decided to archive their original assessment on the lab leak controversy by removing it from their database and revising their assessment as "widely disputed" (see https://www.politifact.com/li-meng-yan-fact-check/)

Such ambiguities inevitably demand data scientists to interpret, standardize or develop new labels from existing data. For instance, to tackle the issue with inconsistent labels in the CLARITI project, the ML model was trained to classify the individual verdicts into three categories: "factual", "misinformation" and "other." The last category was included to handle verdicts which were not conclusive, such as "not sure" –the claims whose verdicts belonged to the "other" category were not included in the final dataset. Training such a model requires labelled verdicts to be standardized even if this introduces new ambiguities and loss of nuance. For example, "Half true" was categorized as "Misinformation". To mitigate these ambiguities, a decision was made in the project to only include claims whose verdicts from the fact-checking organizations were unanimous.

*Training data could also be skewed toward false claims*. While fact-checkers attempt to validate true information and attempt to promote factual content, much of their work is focused on debunking falsehoods.[11] This is reflected in the composition of the ground truth datasets, for example, when they contain disproportionately more samples of falsehoods, or only one label for "fully true" and several ones for dubious content ranging from "mostly true" to "blatantly false." This issue can lead to misrepresentation of truthful content (labelled as such) in datasets which undermines the ability of a model to accurately identify true statements (true negatives) and reduce false positives. The bias toward false claims can be viewed as a technical problem of unbalanced data which developers can attack by attempting to diversify content and sources in the construction of the dataset [28]. However, balancing a dataset is not always a straightforward task. This was true in [project name removed for peer review] where the resulting dataset was largely skewed towards claims belonging to the "misinformation" category (~95% of the claims). A choice was made to not balance the two categories by including, e.g., news articles from "trusted sources", as this would both introduce more bias as well as potentially polluting data from a different data distribution. In attempting to balance the data, ML models could be able to distinguish between new and old data, rather than distinguishing between factual and misleading claims, making the task superficially easier yet futile.

*Datasets bear human selection and cognitive biases*. A crucial and difficult question for the practice of fact-checking is which claims are eligible for assessment. Fact-checkers necessarily incur selection biases when deciding which claims to check and which ones to leave out. This is particularly controversial in the assessment of political discourse where judgements are often passed on statements which may contain a mix of opinion and verifiable facts. According to Uscinsky [71], one of the perils of fact-checking is the choice to assess ideologically charged claims or future predictions for factual accuracy even when these can only be verified retrospectively or are not verifiable at all. Similarly, selection biases might lead to uneven representation of content among fact-checking organizations. For instance, a comparative study of two major fact-checking organizations in the US found that not only did they rarely look at the same selection of statements but even when they did there was little agreement on how they scored ambiguous claims such as "mostly true" or "mostly false" [45]. Selection biases are not only a source of uncertainty, but they normatively influence what types of information are worthy of checking and which narratives are prioritized over others.

## 4.4 Model validation

The merit and utility of a classification model is judged by its ability to predict human generated labels. Once a model is trained, its accuracy can be measured by comparing the resulting classifications against an *unseen* subset of the ground truth data. For example, in the case of models using datasets with labels provided by fact-checkers (e.g., true or false), 100% accuracy on the test set will theoretically equate to the model correctly predicting all the labels given by the fact-checker on data not seen by the model during the training process.

---

[11] Some fact-checking organizations focus exclusively on false and misleading claims (e.g., factcheck.org)

This is a process of internal validation which is typically agnostic to how the model functions in the world and the possibility of downstream harmful impacts. Performance metrics (be they *accuracy*, *precision*, *recall*, and *$F_1$-score*[12]) are commonly used as indicators of relative incremental progress within the field and are used for comparisons against benchmarks of human decision-making or other competing algorithmic techniques. However, these comparisons may be decontextualized; that is, based on metrics alone without regard to the specific (thematic, temporal or cultural) domains in which different models were trained and the qualitative differences between them. Such decontextualization can be misleading as a model trained on e.g., political misinformation, is likely to be inadequate to detect misinformation in the celebrity domain [31]. Since accuracy metrics are not always indicators of good model performance they could be deceiving, particularly in models using imbalanced or unevenly represented datasets which still exhibit relatively high accuracy [73]. In the development of the CLARITI project's dataset, diversity of the data was deemed of high priority, as existing benchmarking datasets are biased towards specific languages, topics or events. As the system sought to detect misinformation within unseen events, the dataset was not merely split at random into a training and testing part. Instead, these splits were created according to distinct events, thus making more consistent evaluations, albeit substantially harder. Further, the dataset was heavily unbalanced (95% of the data belongs to the misinformation category) which means an accuracy metric would not be very telling and therefore $F_1$-scores for the two categories were reported instead.

Despite their salient shortcomings, performance metrics have *performative power*[13] in that they create expectations around, and effectively vouch for, the value of an algorithm. Whether, and how, to deploy an ML model can be informed by various metrics of performance. For instance, if a model exhibits a relatively high level of accuracy in classifying fake content, this can be used as a justification for deploying a system without human moderation. According to Pérez-Rosas et al., [56] models with over 70% accuracy are generally considered as being as good as humans to identify fake news (to use the authors" term), yet they still have considerable room for errors. Metrics of accuracy, precision, recall and $F_1$-score not only provide an opportunity for granular performance evaluation, but they can crucially inform what specific actions can be triggered by a model. For example, a model with high precision (low false positive) and low recall (high false negative) may be deemed more useful in fully automated scenarios as, while it may miss many cases of misinformation, there will be more confidence that those it detects will be correct. On the other hand, in scenarios with human moderation, a model with lower precision, but higher recall, may be more useful as it will retrieve more possible misinformation than the former model, albeit at the expense of false positives, which can be corrected by human moderators.

## 5   Anticipating emergent issues during model deployment

There are several ways in which social media platforms implement detection algorithms. They can either configure hybrid decision-support systems (e.g., AI-assisted fact-checking) or operate as standalone, automated moderation systems with no human intervention. In the case of Meta, content flagged by an algorithm as potentially false is typically relayed to independent fact-checkers who will make decisions on the veracity of the claim [9]. This is a strenuous process requiring a great deal of manual input to process the huge amount of content circulating on social media.

Depending on the platform's policy, detection algorithms can trigger specific corrective actions such as banning/flagging/downranking content or promoting relevant verified information alongside deceitful posts [25,26]. One of the pitfalls of such corrective approaches (and moderation policies at large) is that they are typically applied at global scale (affecting billions of people) with little regard to different demographics and socio-political contexts and in line with the company's (shareholders) values and definitions of what counts as being acceptable.[14] Moreover, there is widespread evidence that major social media platforms have facilitated the formation of knowledge communities where content is circulated and segregated based on shared politics and interests [11,60]. Because of this, platform-wide corrective actions may not only have disparate effects when used across different groups and cultural contexts but pose the risk of reinforcing false beliefs particularly amid those groups where the circulation of falsehoods or conspiracy theories is more prevalent [53]. Existing algorithmic techniques still have limited ability to account for nuances in language, intent, cultural references, or sarcasm [17]. This makes algorithms highly fragile when it comes to "borderline" or tricky cases but also vulnerable to being circumvented by the creators of false content emulating the style of truthful sources or translating posts into other languages. Similarly, the overreliance on seemingly high performing algorithms risks worsening issues of unjustified censorship when content is wrongly identified as being false and is subsequently banned or downranked. Performance metrics are often invoked to frame a complex social problem within a logic of optimization. If errors are low, platforms tend to dismiss them as negligible or outweighed by the benefits of improved efficiency thereby shifting the burden of errors to an acceptable minority of affected users who are faced with appeal processes.[15]

---

[12] Here accuracy is the proportion of the model's predictions which are correct, recall is the proportion of the positive samples which the model correctly predicted, precision is the proportion of the model's positive predictions which are correct. The F1-score is the harmonic mean of the recall and precision, which implies that if one of these two metrics are low then the F1-score will be correspondingly low as well.

[13] The concept of language performativity is used here in the same sense as within language anthropology, gender studies and sociology of expectations. A claim or statement is thought of as performative insofar as it constitutes and *act* which has an effect in the world [see 6,30].

A perhaps more fundamental issue with the use of algorithmic misinformation detection is that it emphasizes the role of individual consumers and producers of misinformation at the expense of downplaying the interests and responsibilities of major technology companies. The business model of social media platforms is based on maximizing the time users spend on their platforms in order to generate advertising revenue. This is achieved through opaque algorithms of personalization and recommendation based on people's behavior, demographics and preferences [7]. The attention economy rewards the circulation of (and engagement with) content regardless of its quality; and in fact, misinformation has been found to consistently receive widespread attention and engagement in social media platforms [19]. The commercial incentive of platforms to maximize engagement is thus at odds with the goal of meaningfully tackling the spread of misinformation and any type of harmful content. Furthermore, there is a risk that automation is positioned as the only solution to the spread of harmful content that ensures business as usual. While we recognize that automated detection can have promising benefits to deal with the scale of misinformation, its development should not foreclose the broader debate around platform regulation and oversight.

# 6 Discussion: Toward a responsible ML agenda

We contend that insofar as a model's outputs are underpinned by a series of contingent assumptions, institutional commitments and socially constructed assessments of facticity, there can be no such thing as an impartial or neutral (mis)information classifier. The contingencies associated with developing ML classification models evidence that multiple reasonable strategies and outcomes are possible and that these are necessarily influenced by the subjectivities, interests and random events implicated in their development. Further, we emphasize that misinformation detection algorithms are highly temporally sensitive: models using historic data may quickly become obsolete and hence need to be routinely assessed considering up-to-date information and changing moderation norms set by platforms and regulatory bodies.

A constructive question arising from the contingencies outlined here is what measures can be taken in the interest of harnessing the social value of algorithmic classification and minimizing any harmful effects. There is no straightforward procedure to establish what the *right* outcomes might look like given that desired outcomes, harmful effects and social preferences might be in conflict. For instance, while some might be in favor of reducing the volume of misinformation online by maximizing a model's true positives with a tolerable error, others will be disproportionally harmed by unfair censorship and undermined freedom of expression resulting from misclassifications. Similarly, some would argue that people have the right to share misinformation particularly if it is harmless, whereas potentially dangerous content could provide a justification for restrictions on freedom of expression. Yet in practice, drawing boundaries between harmful/harmless content and the limits to freedom of expression is seldom a trivial exercise.

---

[14] In the context of the Russian-Ukrainian conflict, for instance, human rights observers flagged the problematic double standards of large platforms when special exceptions were given to expressions related to Ukrainian self-defense while enacting radically different policies in other places like Syria, Yemen and Palestine [3].

[15] As recently admitted by YouTube's representative: "One of the decisions we made [at the beginning of the pandemic] when it came to machines who couldn't be as precise as humans, we were going to err on the side of making sure that our users were protected, even though that might have resulted in a slightly higher number of videos coming down." (Neal Mohan quoted in [1])

These are ongoing tensions which should not be rendered as solvable problems. Instead, the question of how we might produce socially beneficial ("good" or "fair") algorithmic tools calls for careful attention to broader socio-technical, legal, political and epistemic considerations. We suggest developers should endeavor to account for algorithmic contingencies and reflect on the limitations of their creations. This implies a commitment to openness and self-critical reflection, making the assumptions and the various human choices throughout the stages of data curation, model construction and validation available for scrutiny and contestation by external observers and taking their potential for harmful outcomes seriously. While this is an open research challenge, we offer a tentative agenda toward responsible ML classification models.

## 6.1 Reflexivity beyond datasets

Principled and defensible criteria such as relevance, authoritativeness, data structure and timelines of the truth assessments all provide a strong foundation for the curation of ground truth datasets. There are also important ongoing efforts to improve the transparency of datasets which are of relevance here [22–24]. However, we propose that accounting for contingencies, particularly in politically sensitive scenarios, requires going beyond considerations of data accuracy, reliability and quality to acknowledge the complex processes of social construction which configure the development and use of ML models. A recent study by Birhane et al., [5] showed that highly cited ML research has typically ascribed to values of performance, efficiency and novelty over considerations of social needs, harms and limitations; yet most often researchers make implicit allusions to the value neutrality of research. Insofar as developers outsource the assessments of facticity to other actors and select particular topics or events as matters of concern, it becomes more crucial to examine one's own assumptions, biases and commitments which directly influence model development and that these are made available for auditing purposes.

Misinformation classification is by necessity a value-laden practice with profound normative implications concerning the validity, quality, representativeness and legitimacy of knowledge. Linking back to the efforts of feminist scholars in surfacing the politics of knowledge production, we are reminded of the need to reject "view from nowhere" ideals and practice reflexivity [33,67]. In the interpretative research tradition, reflexivity has been a standard of academic rigor and credibility which is attained through acknowledging prior biases, positionalities, experiences and prejudices impacting researchers" claims to knowledge. There is no reason why improving the credibility of scientific endeavors through reflexivity should not extend to the development of machine learning models. Indeed, reflexivity has begun to be invoked in numerous calls for more transparency and accountability in the field of data science at large [15,48,68].

There remain a great deal of practical challenges with attaining the intended virtues of reflexivity in organizational spaces fraught with multiple, conflicting logics such as universities [54]. Despite this, we support a reflexive turn in data science and recommend much needed further research in this direction. At the very least, a reflexive and transparent approach seeks to avoid shifting the blame to the data and external sources, acknowledge partiality (as opposed to deceptive efforts to debias) and the distribution of collective responsibility within the actors and institutions involved in constructing and deploying a model. In order to surface algorithmic contingencies, we suggest transparency reports need to be complemented with reflexive disclaimers about developers" methodological choices, problem statements, institutional affiliations and sources of funding influencing data collection and model construction.

## 6.2 Situated and timely evaluations

Mechanisms should be in place to adjust the behavior of a model or even the decision as to whether to deploy a model or not with regards to changing circumstances, information or situated community norms. For instance, changes in the terms of service of a platform, or relevant local norms and regulations (e.g., GDPR) should be taken into account along with dataset labels changing as a result of new information (e.g., fact checkers changing their original decision). Equally, if a user deletes their post, this should be removed from the training / test set. Thus, datasets, even if they do not collect any more data, do not remain static – in fact they can decrease in size over time – as the training and test data changes, the model performance will change. This would allow for some form of a dynamic and adaptive environment, where published models and their results are continuously re-evaluated. Community benchmarks based on location, language and domain-relevant test sets is one way to encourage this. These evaluations should be consistent so that model comparisons are made with attention to topicality, timing, context, language or different modalities used. Benchmark tests, for instance, could be conducted against models trained on data labelled by different fact-checkers to investigate the impact of potential political, selection or cognitive biases in the outcomes of a model. Further research is needed around how conditioning factors such as fact-checkers' political leanings, domain specialisms and location could be factored into the quantitative or qualitative algorithmic evaluations.

## 6.3 Accounting for and communicating uncertainty

ML researchers have tried to understand and measure uncertainty in different ways [2,35]. Qualitative, also called 'epistemic', uncertainty extends to the lack of knowledge about the outputs of a model or ignorance on the part of the decision maker about the innerworkings of a ML model. This type of uncertainty can be a reflection of several factors that are baked into a model such as subconscious biases, inaccuracies or gaps in the data, as well as discretionary forms of data reduction or standardization of labels that are sometimes carried out by developers. For instance, while the temporary fix of omitting ambiguous verdicts (such as "half true" or "mostly false") might reduce the burden for moderators and (superficially) increase accuracy, it comes at the cost of unwanted outcomes such as casting doubt on legitimate information, doing away with important nuances in language, or leaving subtle misleading claims unaddressed.

Model construction is also impacted by aleatoric uncertainty. The data collection process from a social media platform can be stochastic for various reasons –Twitter for example provides only a sample of the stream of online posts for searches (thus two people searching for "coronavirus" using that API may collect different results and thus create a different dataset). Even if not using this, arbitrary decisions (such as only keeping a subset of the available data) made throughout the data collection process are often not completely documented and may thus be irreproducible. Thus, the data collection systems themselves should be made public and available. Even if not fully reproducible (due to the dynamic nature of the social media platform, or their stochastic APIs), developers should provide complete executable documentation on the data collection process. This approach is considered as a way forward in the CLARITI project: the data collection platform has been released on GitHub [51] so others can see and execute the exact code used to build the dataset, and thus all decisions that were made.

Measuring and reporting the uncertainty of a model can be crucial in aiding human intervention and increasing the transparency of the system [2]. Measures of uncertainty could be published in tandem with other metrics as part of the model evaluation. This could help to avoid overreliance on algorithms and minimize ambiguous outcomes by helping human moderators but also to ponder alternative interventions such as adding context to ambiguous content or links to contrasting news.

## 7 Conclusion

Through an interdisciplinary collaboration, we have critically examined the efforts to identify and manage online misinformation at scale using machine learning classification models and have proposed an agenda for a more reflexive and responsible development of these tools. The development and widespread use of automated misinformation detection systems raise pressing political, epistemic and ethical issues. We argue that, albeit promissory developments, these tools are highly contingent to the epistemic status of their ground truths, and the assumptions, choices and definitions underpinning their development, the contexts within which they are deployed and the interests of powerful actors in vetting the circulation of information online.

We laid out a series of contingencies across the different stages in the construction of these models and assessed how assumptions of expertise and legitimacy, ideological biases, and commercial and (geo)political interests may influence the normative outcomes of models which are predicated as robust, accurate and high performing. We note that while our analysis is grounded on a specific issue tackled by ML, similar concerns are likely to hold true in other areas, particularly in the moderation of hate speech and terrorist content. This marks paths of future inquiry where our analytical approach could be used to interrogate the epistemologies and forces driving other algorithmic systems.

This study exemplifies an attempt at bringing social sciences methods and theory into conversation with those of data science. We hope our contribution sparks further exploration of how data scientists and social scientists can work together so as to break with traditional and perhaps unproductive divisions of labor when research questions seem to fall out of the remit of one discipline or the other [49,65]. In considering the wealth of literature on RI, STS, critical algorithm studies and AI ethics, we also recognize that previous learning cannot be taken as readily intelligible and applicable to those expected to apply said learning and therefore that in-situ inquiries are imperative. We thus approached this study in an open-ended and co-productive fashion, conducting iterative cycles of observation, interpretation, validation and calibration among the research team. We call for further experimentation of this type that supports the embedding of critical scholarship within data and computational sciences.

## Disclosure Statement

## Ethics declaration

This study was designed as an internal research project and approved by the ethics committee of the authors' institution following fair data management practices, informed consent and responsible research and innovation considerations.

## Funding

This research was supported by REPHRAIN: The National Research Centre on Privacy, Harm Reduction and Adversarial Influence Online, under UKRI grant: EP/V011189/1.